\documentclass[11pt,a4paper,dvipsnames]{article}
\usepackage{authblk}
\usepackage{times}
\usepackage{latexsym}
\usepackage{url}

\usepackage{geometry}
\usepackage{bm,bbm}
\usepackage{amsmath,amsthm,amssymb}
\usepackage{float}								
\usepackage{xcolor}
\definecolor{cRed}{HTML}{DA5527}
\definecolor{cGold}{HTML}{EEB11D}
\definecolor{cBlue}{HTML}{0A73B9}
\definecolor{cGreen}{HTML}{008D0A}

\usepackage{tikz}
\usetikzlibrary{positioning}
\usetikzlibrary{shapes.geometric}

\usepackage{subfig}
\usepackage{graphicx}
\usepackage{booktabs}							
\usepackage{array}								    
\usepackage{enumitem}                           
\setlist{noitemsep, topsep=0pt, leftmargin=*}   
\theoremstyle{plain}

\newtheorem{prop}{Proposition}

\theoremstyle{definition}
\newtheorem{defn}{Definition}

\newtheorem*{rem*}{Remark}

\def\r{\mathbf{r}}
\def\s{\mathbf{s}}
\def\t{\mathbf{t}}

\def\x{\mathbf{x}}

\def\z{\mathbf{z}}

\def\A{\mathbf{A}}

\def\N{\mathbf{N}}

\def\bfeta{\bm{\eta}}

\def\bmu{\bm{\mu}}

\def\brho{\bm{\rho}}
\def\btheta{\bm{\theta}}

\def\bTheta{\bm{\Theta}}

\def\cH{\mathcal{H}}
\def\cI{\mathcal{I}}

\def\cL{\mathcal{L}}

\def\cZ{\mathcal{Z}}

\def\bbE{\mathbb{E}}
\def\bbI{\mathbb{I}}
\def\bb1{\mathbb{1}}

\def\defas{\triangleq}

\def\Multi{\mathrm{Multi}}

\def\Gam{\mathrm{Gamma}}

\def\Dir{\mathrm{Dir}}

\usepackage[draft]{changes}             
\definechangesauthor[name={Wei}, color=blue]{WZ}
\definechangesauthor[name={Fan}, color=red]{FB}

\allowdisplaybreaks

\newcommand{\reducemargin}{\vspace{-0.1in}}

\newcommand{\myparagraph}[1]{\reducemargin \paragraph{#1}}

\usepackage[round]{natbib}
\begin{document}

\title{Who Started It? \\ Identifying Root Sources in Textual Conversation Threads}


 \author[*]{\bf Wei Zhang}
 \author[$\dag$]{\bf Fan Bu}
 \author[$\dag$]{\bf Derek Owens-Oas}
 \author[$\dag$]{\bf Katherine Heller}
 \author[*]{\bf Xiaojin Zhu}
 \affil[*]{Department of Computer Science, University of Wisconsin-Madison}
 \affil[$\dag$]{Department of Statistical Science, Duke University}
 
\date{}

\maketitle

\begin{abstract}

In textual conversation threads, as found on many popular social media platforms, each particular user text comment either originates a new thread of discussion, or replies to a previous comment. An individual who makes an original comment
---termed as the ``root source''---is a topic initiator or even an information source, and identifying such individuals is of particular interest. The reply structure of comments is not always available (e.g. in the proliferation of a news event), and thus identifying root sources is a nontrivial task. In this paper, we develop a generative model based on marked multivariate Hawkes processes, and introduce a novel concept, \emph{root source probability}, to quantify the uncertainty in attributing possible root sources to each comment. A dynamic-programming-based algorithm is then derived to efficiently compute root source probabilities. Experiments on synthetic and real-world data show that our method identifies root sources that match ground truth and human intuition. 
\end{abstract}

\section{Introduction}
\label{sec:intro}
Textual conversation threads---individual textual utterances made sequentially by a group involved in discussion, in a conversational manner---are widely observed on social media, in online forums, and in daily conferences and debates. Every textual comment can be thought of as either original or a response, where an original comment may prompt a new branch of discussion in the conversation. 
The individual responsible for an original comment is thus a \emph{topic initiator} who potentially shifts the direction of discussion, or even an \emph{information source} for the community. On the Internet, a term ``original poster (OP)'' has been developed for 
those ``topic-initiating'' individuals, 
in order to attribute credit when novel ideas or meaningful talking points are brought up, or to lay blames when misinformation is spread. In a news-related user community on Twitter, for example, a journalist exposing the Facebook-Cambridge Analytica data scandal may be the OP for many follow-up posts and discussions, whereas the National Rifle Association account might be an OP of a series of tweets regarding gun violence. 

In this paper, we use the term ``\emph{root source}'' instead of ``OP'' to accommodate general textual conversation threads, 
and tackle the task of 
identifying possible root sources for each comment. 
Learning root sources is of great interest to a wide audience including sociologists, psychologists and policy makers, and helps solve problems such as credit attribution, rumor tracing, and social power inference \cite{danescu2012echoes}. 

Identifying root sources in general conversation threads, however, presents two major challenges. 
First, although one can pin down the root sources by back-tracing the \emph{direct reply structure} to the earliest posts for cascades on online platforms (e.g., Twitter),
such clear structure, unfortunately, is almost never available in offline textual conversation threads (e.g., court room transcriptions). Thus, one has to infer the hidden reply structure in order to identify root sources. 
Second, even though each comment only corresponds to one root comment and thus one root source, there could exist multiple root sources during the whole conversation, as individuals may intermittently start new topics, attempting to alter and lead the direction of discussion. 

We propose to achieve the goal of root source identification in textual conversation threads using multivariate Hawkes processes (MHPs) with textual marks, a class of mutually-exciting point processes. 
This model leverages \textbf{three} key aspects of textual conversation threads:
(a) \emph{temporal locality}--individuals tend to respond to recent comments; (b) \emph{individual heterogeneity}--different individuals can have drastically different intrinsic comment rates and tendencies to reply; (c) \emph{vocabulary inheritance}--an individual tends to adopt certain vocabulary words from those whom he/she is responding to. 
To quantify the uncertainty of identified root sources for each comment, we introduce a novel concept, \emph{root source probability}, and develop an algorithm to compute root source probabilities for all comments efficiently. This allows us to, for example, identify an individual information generator in conversation with high probability.

\vspace{0.05in}
\myparagraph{Main Contributions:}


\begin{itemize}
  \item A generative model based on marked MHPs is proposed. The model captures the three listed essential aspects of textual conversation threads: temporal locality, individual heterogeneity, and vocabulary inheritance. 
  
  \item A novel concept, \emph{root source probability}, is introduced. It quantifies the posterior uncertainty in identifying root sources. 
  
  \item An efficient, dynamic-programming algorithm is derived to compute root source probabilities. 
  
\end{itemize}

\vspace{0.05in}
The rest of the paper is organized as follows. 
Necessary background is provided in Sec.\,\ref{sec:background}, and the parameterization of the marked MHPs model is described in Sec.\,\ref{sec:model}. 
Sec.\,\ref{sec:theory} formally defines root source probability and derives an efficient computation algorithm.
Sec.\,\ref{sec:learning} explains the parameter estimation procedure for the model, and Sec.\,\ref{sec:related_work} discusses related work. 
Finally, experiments are presented in Sec.\,\ref{sec:experiments}, followed by conclusions in Sec.\,\ref{sec:conclusion}. 


\section{Background}
\label{sec:background}

\subsection{Marked Multivariate Hawkes Processes}

An $S$-dimensional multivariate Hawkes process (MHP)  \cite{hawkes1971point, hawkes1971biometrika, embrechts2011multivariate} is a coupling of $S$ counting processes $\N(t) \defas \left[ N^{(s)}(t) \right]_{s \in [S]}$, each of which counts up the number of events occurring on source $s$ before time $t$. 
We use $[n]$ as a shorthand for the set $\{1, \ldots, n\}$ for any positive integer $n$ and $\defas$ for ``is defined as''.  
A sample of an MHP is a sequence of events $e_{1}, e_2, \ldots$, where the $i$-th event $e_i \defas (t_i, s_i)$ consists of a timestamp $t_i$ and a dimension/source label $s_i$, indicating when and from which source the event occurs. 
The history, $\cH_{t-}=\{e_i: t_i < t\}$, is the set of events that occur \emph{strictly before} $t$, for any $t > 0$. The \emph{conditional intensity} $\lambda^{(s)}(t|\cH_{t-})$ for the $s$-th process of the MHP, for $s \in [S]$, takes the form 
\begin{equation}
\lambda^{(s)}(t|\cH_{t-}) \defas \mu^{(s)}(t) + \sum_{t_i<t} \lambda_i^{(s)}(t),
\label{eq:lambda_i(t)}
\end{equation}
where $\mu^{(s)}(\cdot)$ and $\lambda_i^{(s)}(\cdot)$ are the base intensity and the excited intensity of source $s$ attributed to the previous event $e_i$, respectively.

An important extension of MHPs is marked MHPs, which introduce to each event $e_i$ a mark $\x_i$. 
It is often assumed that mark $\x_i$ is drawn, conditioned on  $t_i$ and $s_i$,  from a \emph{mark density} $P(\cdot|t_i,s_i,\cH_{t_i-})$.
As the notation suggests, in the most general case the mark density may depend on  timestamp $t_i$ and  source label $s_i$ of the $i$-th event, as well as all historical events before $t_i$, i.e. $H_{t_i-}$.

\subsection{Branching Structure}
\label{subsec:branching-structure}

An equivalent view of MHPs is Poisson clustering processes \cite{Rasmussen:2011}. The Poisson clustering processes start with $S$ inhomogeneous Poisson processes (IPPs), each of which is associated with a base intensity $\mu^{(s)}(t)$ and forms its own cluster. 
Then the IPPs corresponding to the base intensities begin to generate events, which are called \emph{immigrants}.   
Whenever an immigrant $e_i$ is generated, it adds to each cluster $s$ a new IPP with intensity $\lambda_i^{(s)}(\cdot)$, which further generates the so-called \emph{offsprings}. 
Eventually, the sample comprises all the events---both immigrants and offsprings---from all the clusters.

This Poisson clustering point of view introduces a latent \emph{branching structure}, defined by the \emph{parental relationship} between events in the sample. 
Specifically, let the one-hot vector $\z_i \defas \left[ z_{ij} \right]_{j=0}^{i - 1} \in \{0, 1\}^{i}$  be the parent variable for event $e_i$, such that $z_{i0}=1$ if $e_i$ is an immigrant from $\mu^{(s_i)}(\cdot)$, and $z_{ij}=1$ if $e_i$ is an offspring from $\lambda_j^{(s_i)}(\cdot)$. 
Based on the superposition property of Poisson processes, the distribution of $\z_i$ conditioned on timestamp $t_i$, source label $s_i$, and the history $\cH_{t_i-}$ is
\begin{equation}
\begin{small}
P(\z_i|t_i,s_i,\cH_{t_i-}) =
\begin{cases}
\frac{\mu^{(s_i)}(t_i)}{\lambda^{(s_i)}(t_i|\cH_{t_i-})} & z_{i0}=1, \\
\frac{\lambda_j^{(s_i)}(t_i)}{\lambda^{(s_i)}(t_i | \cH_{t_i-})} & z_{ij} = 1, \\
0 & \text{o.w.}
\end{cases}
\end{small}
\label{eq:P(z_e)}
\end{equation}

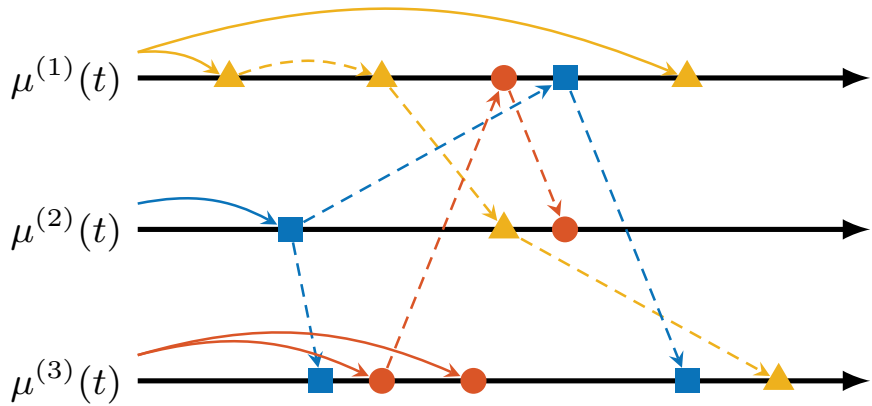
\begin{figure}[tbp]
\centering
\resizebox{0.8\columnwidth}{!}{
\footnotesize

\begin{tikzpicture}[
	scale=2.5,
    axis/.style={very thick, ->, -latex},
    immigrant/.style={-stealth, semithick},
    offspring/.style={densely dashed, -stealth, semithick},
    dot/.style={circle, minimum size=1.5ex, inner sep=0, fill=cRed},
    triangle/.style ={regular polygon, regular polygon sides=3, minimum size=2.1ex, inner sep=0, fill=cGold},
    square/.style ={regular polygon, regular polygon sides=4, minimum size=2ex, inner sep=0, fill=cBlue},
    ]
    
    \draw [axis] (0, 1) node[left] (s1) {$\mu^{(1)}(t)$} -- (2.4, 1);
    \draw [axis] (0, 0.5) node[left] (s2) {$\mu^{(2)}(t)$} -- (2.4, 0.5);
	\draw[axis] (0, 0) node[left] (s3) {$\mu^{(3)}(t)$} -- (2.4, 0);

    \coordinate[triangle](e11) at (0.3, 1); 
    \coordinate[triangle](e12) at (0.8, 1); 
    \coordinate[dot](e13) at (1.2, 1); 
    \coordinate[square](e14) at (1.4, 1);
    \coordinate[triangle](e15) at (1.8, 1);
    
    \coordinate[square](e21) at (0.5, 0.5); 
    \coordinate[triangle](e22) at (1.2, 0.5); 
    \coordinate[dot](e23) at (1.4, 0.5); 
    
    \coordinate[square](e31) at (0.6, 0); 
    \coordinate[dot](e32) at (0.8, 0); 
    \coordinate[dot](e33) at (1.1, 0); 
    \coordinate[square](e34) at (1.8, 0);
    \coordinate[triangle](e35) at (2.1, 0); 
    
    \draw [immigrant,color=cGold] (s1)  to [bend left=20] (e11);
    \draw [offspring,color=cGold] (e11) to [bend left=20] (e12);
    \draw [offspring,color=cGold] (e12) -- (e22);
    \draw [immigrant,color=cGold] (s1)  to [bend left=20] (e15);
    \draw [offspring,color=cGold] (e22) -- (e35);

    \draw [immigrant, color=cBlue] (s2)  to [bend left=20] (e21);
    \draw [offspring, color=cBlue] (e21) -- (e14);
    \draw [offspring, color=cBlue] (e21) -- (e31);
	\draw [offspring, color=cBlue] (e14) --	 (e34);

    \draw [offspring, color=cRed] (e13) -- (e23);
    \draw [immigrant, color=cRed] (s3)  to [bend left=20] (e32);
    \draw [immigrant, color=cRed] (s3)  to [bend left=20] (e33);
    \draw [offspring, color=cRed] (e32) -- (e13);

\end{tikzpicture}
}
\caption{A branching structure for a MHP.  Immigrants and offsprings are pointed by solid and dashed arrows, respectively.  Events sharing the same root source are represented with the same marker. For example, all red circles  trace back to and thus share the same root source---source 3---even though they are made by different sources. }
\label{fig:branching-structure}
\reducemargin
\reducemargin
\end{figure}

An important property of the branching structure is that an MHP can be viewed as a forest with $S$ trees. 
This is because the ``parent'' of each event $e_i$ is either an earlier event $e_j$ or the base intensity $\mu^{(s)}$ (i.e. $e_i$ originates directly from source $s$). 
Therefore, given a branching structure, each event can be traced back to its ``root'', the source from which it originates, as illustrated in Figure~\ref{fig:branching-structure}.

Given the nature of the branching structure, a more definitive form for the mark density $P(\cdot|t_i,s_i,\cH_{t_i-})$ is often considered, in order to emphasize a direct dependency of the mark of an event on its parent event \cite{Rasmussen:2011}.
Conditioned on the parent variable $\z_i$,  the mark $\x_i$ of each event $e_i$ is assumed to be drawn from 
\begin{equation}
\begin{small}
  P(\cdot|t_i,s_i,\z_i,\cH_{t_i-}) \defas
  \begin{cases}
  f(\cdot|t_i,s_i)     & z_{i0}=1, \\
  f(\cdot|t_i,s_i,e_j) & z_{ij}=1,
  \end{cases}
\end{small}
\label{eq:conditioned-mark-density}
\end{equation}
where $f(\cdot|t_i,s_i)$ and $f(\cdot|t_i,s_i,e_j)$ are two parameterized probability densities. Combining \eqref{eq:P(z_e)} and \eqref{eq:conditioned-mark-density} and marginalizing out the parent variable $\z_i$ lead to   
\begin{equation}
\begin{small}
\begin{aligned}
P(& \x_i | t_i,s_i, \cH_{t_i-}) 
= \frac{\mu^{(s_i)}(t_i)}{\lambda^{(s_i)}(t_i|\cH_{t_i-})}f(\x_i|t_i,s_i)  \\
& + \sum_{j<i}   \frac{\lambda_j^{(s_i)}(t_i)}{\lambda^{(s_i)}(t_i|\cH_{t_i-})} f(\x_i|t_i,s_i,e_j),
\end{aligned}
\end{small}
\label{eq:mark-density}
\end{equation}
which implies that this mark density is in fact a mixture with weights proportional to the different intensity components.

Unless stated otherwise, in the remaining paper we shall focus on the marked MHP, of which the conditional intensities and mark density are of the forms specified in \eqref{eq:lambda_i(t)} and \eqref{eq:mark-density}, respectively. 
%

\section{Modeling Conversational Textual Cascades with Marked MHPs}
\label{sec:model}
In this section, we propose a specific parameterization of marked MHPs to characterize the generative process of textual conversation threads. 

Suppose there are $S$ individuals participating in a conversation. We denote the $i$-th comment by $e_i \defas (t_i, s_i, \x_i)$, indicating that it is made by individual $s_i$ at time $t_i$ with textual content $\x_i$.  
We represent the content $\x_i$ as a bag-of-words vector for a vocabulary of $V$ tokens.
Let $n$ be the total number of comments observed in time window $[0, T]$, and define $\cH_T \defas \{e_i: i \in [n]\}$. 
Note that, in the rest of the paper, depending on the context, the words ``event'' and ``comment'', and ``source'' and ``individual'' may be used interchangeably. 

There are \textbf{four} components of a marked MHP that require a concrete parameterization: 
base intensity $\mu^{(s)}(\cdot)$, exciting intensity $\lambda_i^{(s)}(\cdot)$, mark density for immigrants $f(\cdot| t_i, s_i)$, and mark density for offsprings $f(\cdot | t_i, s_i, e_i) $. 
Furthermore, it is desired that the parameterization reflects the key aspects (discussed in Sec. \ref{sec:intro}) of textual conversation threads-- \emph{temporal locality}, \emph{individual heterogeneity}, and \emph{vocabulary inheritance}. 

We adopt the following factorized forms for $\mu^{(s)}(\cdot)$ and $\lambda_i^{(s)}(\cdot)$:
\begin{align}
\mu^{(s)}(t) & \defas\rho_s\bar{\mu}^{(s)}(t), \\
\lambda_i^{(s)}(t) & \defas \alpha_{s,s_i}\beta(\x_i)\kappa^{(s)}(t_i,t). 
\end{align}
Here $\rho_s$ and $\bar{\mu}^{(s)}(\cdot)$ are the multiplier and shape function of the base intensity of source $s$, respectively; 
the matrix $\A \defas \left[ \alpha_{ss'} \right]_{ss'}$ characterizes the strength of mutual excitation between sources;
the function $\beta(\cdot)$ quantifies the impact of different textual contents; 
and $\kappa^{(s)}(\cdot, \cdot)$'s are normalized \emph{decay} kernels such that   $\int_{t}^{\infty}\kappa^{(s)} (t,t')dt' = 1$ for $s \in [S]$ and $t > 0$. 
Individual heterogeneity is reflected by the source-specific choice of $\brho$ and $\A$, as well as $\bar{\bmu} (\cdot)$ and $\bm{\kappa}(\cdot)$, and temporal locality is reflected by the decaying property of the kernels $\bm{\kappa}(\cdot)$. 

Next, we specify the mark densities based on the multinomial distribution and its mixture.  
Specifically, let $\Multi(m, \btheta)$ be the multinomial distribution parameterized by $m$ and $\btheta$. 
Conditioned on the text length $L_i$ for event $e_i$, we consider the mark densities
\begin{align*}
f(\cdot | t_i, s_i) & \defas \Multi(L_i, \btheta^{(s_i)}), \\
f(\cdot | t_i,s_i,e_j) & \defas \Multi(L_i, (1 - \gamma) \btheta^{(s_i)} + \gamma\tilde{\x}_j),
\end{align*}
where $\gamma \in (0, 1)$ is a scalar parameter, and $\tilde{\x} \defas \x / \sum_{v \in [V]}x_{v}$ is the normalized bag-of-words vector for any $\x$. 
Note that $f(\cdot |t_i,s_i,e_j)$ is equivalent to a word-level multinomial mixture; that is, each token made by individual $s_i$ is drawn i.i.d. from $\Multi(1, \btheta^{(s_i)})$  with probability $1-\gamma$  and from  $\Multi(1, \tilde{\x}_j)$ with probability $\gamma$. Thus, vocabulary inheritance is captured through this \emph{word-level mixture} design.  

The graphical model of the above parameterization is included in Appendix \ref{ap:subsec:graphical-model}.

\section{Root Source Probability}
\label{sec:theory}
\subsection{Overview}
\label{subsec:root-proba-overview}
The parental relationship for events of marked MHPs is analogous to the direct replying relationship for comments. From the viewpoint of the branching structure (as in Sec.\,2.2), a comment that initiates a new topic is an immigrant event, and one that replies to existing comments is an offspring event. 
Therefore, inference about a marked MHP translates into identifying root sources in textual conversation threads.  
This motivates us to consider the following (informally defined for now) quantity for each event $e_i$:
\begin{equation}
Pr(\text{$s$ is the root source of $e_i $} | \cH_T),
\label{eq:rooted-probability-informal}
\end{equation}
which is in the form of a probability to address the uncertainty of root source identification.
We name this novel quantity \emph{root source probability}.

We propose to tackle the task of identifying root sources with a \emph{two-stage approach}: 
first, estimate the parameters of the $S$-dimensional marked MHP model specified in Sec.\,\ref{sec:model}; second, apply the learned model to compute the root source probability in  \eqref{eq:rooted-probability-informal}~for each event $e_i$ and source $s$. 
\begin{table}[htbp]
  \centering
  \resizebox{0.9\columnwidth}{!}{
    \begin{tabular}{ll}
\toprule
\textbf{Notation} & \textbf{Description}\\
\midrule
$S$         & the number of dimension/sources \\
$\N_s(t)$   & the counting process of the $s$-th dimension.  \\
$e_i = (t_i, s_i, \x_i)$       & $i$-th event \\
$t$         & time \\
$\mu(\cdot)$    & base intensity \\
$\lambda(\cdot)$ & intensity \\
$\z_{\cI}$  & branch structure \\
$\delta_{\cdot, \cdot}$ &  Kronecker delta function  \\
$\r_i$ & root source probability \\ 
$r^{(s)}_i$    & $s$-root probability \\
$V$    & vocabulary size \\
$\A$    & influence matrix \\
$\gamma$    & word inheritance rate  \\
$\btheta^{(s)}$    & vocabulary parameter  \\
$a_\rho^{(s)}, b_\rho$    & Bayesian prior parameter for base rate  \\
$a_\alpha^{(s)}, b_\alpha$    & Bayesian prior parameter for infectivity  \\
\bottomrule
\end{tabular}

  }
  \caption{Table of notation and parameters.}
  \label{tab:notation}
  \reducemargin
\end{table}

The first stage, parameter estimation, will be explained in Sec.\,\ref{sec:learning}.
We shall describe the second stage in the next subsection by formally defining the root source probability and then stating an efficient procedure to compute it. To facilitate narration, we provide a list of notation in Table \ref{tab:notation}.

\subsection{Definition and Computation}
\label{subsec:root-proba-details}

Suppose that a sample $\cH_T$ of $n$ events is observed for a marked MHP in time window $[0,T]$. 
Let $\cI\subseteq[n]$ be an index set over events in $\cH_T$, and $\z_{\cI} \defas [\z_i]_{i\in\cI}$ be the collection of parent variables which collectively define the  branching structure over all events $[e_i]_{i \in \cI} $. 
Define $\cZ(\cI)$ as the space of all possible $\z_{\cI}$'s.
For event index $i \in \cI$, let $\cZ_i^{(1)},\cZ_i^{(2)},\ldots,\cZ_i^{(S)}$ be a partition for $\cZ(\cI)$ such that  
\begin{align*}
    \cZ_i^{(s)}(\cI)\defas \{ & \z_{\cI} \in \cZ(\cI) \, |\, \text{$s$ is the root source} \\
             & \text{ of $e_i$ according to } \z_{\cI} \}.
\end{align*}

We define the root source probability as follows. 

\begin{defn}
Given an $S$-dimensional (marked) multivariate Hawkes process with sample $\cH_T$ of $n$ events, the \emph{ $s$-root probability} of event $e_i$ is defined as
\begin{equation}
r_i^{(s)} \defas \sum_{\z_{[n]}\in\cZ_i^{(s)}([n])}P(\z_{[n]}|\cH_T),
\label{eq:rooted-probability-definition}
\end{equation}
and $\r_i \defas  [r_i^{(s)}]_{s \in [S]}$ is called the \emph{root source probability} of event $e_i$.
\end{defn}

This definition seems to indicate that the computation of root source probability is intractable even for a \emph{single} event, as it requires summing over all the posterior probabilities of all branching structures in $\cZ_i^{(s)}([n])$, the size of which grows factorially with the number of events $n$. However, it is actually feasible to carry out efficient computation for \emph{all} events, given certain independence properties
of the marked MHPs:

Let $\cH_t \defas \{e_i : t_i \leq t \}$ be the historical events \emph{up to}\footnote{This notation $\cH_{t}$ can include events occurring at $t$,  slightly different from $\cH_{t-}$.} timestamp $t$ for any $t > 0$. Then,
\begin{enumerate}
\item All parent variables $[\z_i]_{i \in [n]}$ are mutually independent conditioned on $\cH_T$;
\item The parent variables of the existing events $\z_{[i]}$ and the future events $\cH_T\backslash\cH_{t_i}$ are independent conditioned on $\cH_{t_i}$ for any $i\in [n]$. 
\end{enumerate}
Both properties can be easily verified from the generative procedure of the marked MHP  described in Sec.\,\ref{subsec:branching-structure}. They further imply that $P(\z_{[i]}|\cH_{t})=P(\z_{[i]}|\cH_{t_i})$, for any $i\in[n]$ and $t\geq t_i$. 

Furthermore, the following proposition shows that root source probabilities for all events in $\cH_T$ can be computed recursively. 
\begin{prop}
\label{prop:rooted-probability-calculation}
Given an $S$-dimensional marked MHP with a sample $\cH_T$ of  $n$ events, the root source probability $\r_i$ for any $i\in[n]$ satisfies 
\begin{align}
r_i^{(s)} \propto &~\delta_{s_i,s} \mu^{(s)}(t_i) f(\x_i|t_i,s_i) \nonumber \\
& + \sum_{j<i}r_j^{(s)} \lambda_j^{(s_i)}(t_i) f(\x_i|t_i,s_i,e_j),
\label{eq:rooted-probability-calculation}
\end{align}
where $\delta_{x, y} \defas \bbI(x = y)$ is the Kronecker delta function. 
\end{prop}

Thus, the root source probabilities for all events in $\cH_T$ can be efficiently computed via dynamic programming. The proof of this proposition is in Appendix \ref{proof:rooted-probability-calculation}.

\section{Parameter Estimation}
\label{sec:learning}
The parameters for the model described in Sec.\ref{sec:model} are $\bTheta \defas \{\brho,\A,\btheta,\gamma\} $.
Given the observed event sequence $\cH_T$,
parameters are estimated through maximizing the marginal log-likelihood 
\begin{align*}
\cL(\bTheta) & \defas  \log  P(\cH_T | \bTheta) \\
& = \log \Big( \sum_{\z_{[n]} \in \cZ([n])} P(\cH_T, \z_{[n]}|\bTheta) \Big).
\end{align*}
Evaluation of $\cL(\bTheta)$, however, is intractable since the size of $\cZ([n])$ grows factorially with the number of events $n$. 

To address this issue, we adopt the variational expectation-maximization (EM) method \cite{bernardo2003variational}, following prior works \cite{He2015,Yang2013,hoffman2013stochastic}.
The main idea of variational EM is to approximate the posterior distribution $P(\z_{[n]} | \cH_T, \bTheta)$ with a proposed distribution $Q$, construct a lower bound surrogate $\tilde{\cL}$ for $\cL$, and maximize $\tilde{\cL}$ over $\bTheta$ and $Q$ alternatively.

The complete likelihood for an event sequence $\cH_T$ and branching structure $\z_{[n]}$ is \cite{Rasmussen:2011}
\begin{align*}
  P(\cH_T, & \z_{[n]}|\bTheta)=  \exp\left[-\int_{0}^T\sum_s\lambda^{(s)}(t|\cH_{t-})\,dt \right] \\
  & \times \prod_{i=1}^n\left[\mu^{(s_i)}(t_i)f(\x_i|t_i,s_i)\right]^{z_{i0}}  \\
  & \times \prod_{i=1}^n\prod_{j<i}\left[\lambda_j^{(s_i)}(t_i)f(\x_i|t_i,s_i,e_j) \right]^{ z_{ij} }. 
\end{align*}
We adopt the mean-field variational approach and choose the proposed distribution $Q$ to be the fully factorized multinomial, 
\begin{equation}
    Q(\z_{[n]}) \defas \prod_{i=1}^n \Multi(\z_i|1, \bm{\eta}_i),
\label{eq:Q}
\end{equation}
where $\bfeta_i \in [0, 1]^{i}$ is the parameter for the $i$-th multinomial. 
We then construct a lower-bound surrogate for $\cL(\bTheta)$ using the evidence lower bound:
\begin{equation}
\begin{aligned}
\cL(\bTheta) \geq  & \bbE_{Q}[\log P(\cH_T,\z_{[n]}|\bTheta)] \\
& - \bbE_{Q}[\log Q(\z_{[n]})]\nonumber \defas  \tilde{\cL}(\bTheta, \bfeta).
\end{aligned}    
\end{equation}
Since $\tilde{\cL}(\Theta, \bfeta)$ has a tractable, closed form (see Appendix \ref{ap:subsec:ELBO}), we are able to estimate $\bTheta$ by solving the following optimization problem: 
\begin{equation}
    \max_{\bTheta, \bfeta} \tilde{\cL} (\bTheta, \bfeta).
\label{eq:VEM-objective}
\end{equation}
We maximize \eqref{eq:VEM-objective} by block-coordinate ascent with the following updates for each parameter block. 

\myparagraph{Update $\bfeta$}

Maximizing \eqref{eq:VEM-objective} with respect to $\bfeta_i$'s  leads to the following closed-form updates: for $i \in [n]$ and $j  \in [i - 1] $,
\begin{align*}
    \eta_{i0} & \propto \mu^{(s_i)}(t_i)f(\x_i|t_i,s_i), \\
    \eta_{ij} & \propto \lambda_j^{(s_i)}(t_i)f(\x_i|t_i,s_i,e_j).
\end{align*}

\myparagraph{Update $\brho$ and $\A$ with Empirical Bayes}

When prior knowledge about unknown parameters is available, reference priors are often adopted to improve model performance \cite{robbins1964empirical}. 
We adopt independent Gamma priors on $\brho$ and $\A$, i.e., $\rho_s \sim \Gam(a_{\rho}^{(s)}, b_{\rho})$ and $\alpha_{s,s'} \sim \Gam(a_{\alpha}^{(s)},b_{\alpha})$. 
Combining such priors and maximizing the likelihood surrogate $\tilde{\cL}(\bTheta, \bfeta)$ with respect to $\rho_s$ and $\alpha_{s,s'}$, we obtain the following updates (see Appendix \ref{ap:subsec:update-details-EB} for details):
\begin{align*}
\rho_s & =\frac{a_{\rho}^{(s)} - 1 + \sum_{i=1}^n \delta_{s_i,s}\eta_{i0}}{b_{\rho}+\int_{0}^T\bar{\mu}^{(s)}(t)\,dt}, \\
\alpha_{s,s'} & = \frac{a_{\alpha}^{(s)} - 1 +\sum_{i=1}^n\sum_{j<i}\delta_{s_i,s} \delta_{s_j,s'}\eta_{ij} }
                      {b_{\alpha} + \sum_{i=1}^n \delta_{s_i, s'}\beta(\x_i)\int_{t_i}^T\kappa^{(s_i)}(t_i,t)\,dt }  .
\end{align*}

\myparagraph{Update $\btheta$ and $\gamma$}

To obtain the updates for $\btheta$ and $\gamma$, one needs to solve the following sub-optimization problem:
\begin{equation} 
\begin{aligned}
\max_{\btheta,\gamma} & \sum_{s=1}^{S}\sum_{i=1}^n\sum_{v=1}^{V}\delta_{s_i,s}g_{i,v}^{(s)}(\btheta,\gamma), \\
& \text{s.t.} \sum_{v=1}^{V}\theta_{v}^{(s)}=1, \forall s \in [S], 
\end{aligned}
\label{eq:update-obj-theta-gamma}
\end{equation}
where
\begin{equation}
\begin{aligned}
    g_{i,v}^{(s)} & (\btheta,\gamma) \defas \eta_{i0} x_{i,v} \log\theta_{v}^{(s)} \\
    & + \sum_{j<i}\eta_{ij}x_{i,v} \log \left[(1-\gamma)\theta_{v}^{(s)}+\gamma\tilde{x}_{j,v}\right].
\end{aligned}
\label{eq:g_siv}
\end{equation}

We only sketch the optimization strategy for \eqref{eq:update-obj-theta-gamma} here and include more details in Appendix \ref{ap:subsec:update-details-vocabulary-param}. Apply Jensen's Inequality with a coefficient $\xi^{(s)}_{j,v} \in (0, 1)$ to each logarithm term in the summation of \eqref{eq:g_siv}; this allows us to construct $\tilde{g}_{i,v}^{(s)}$, a lower bound for each $g_{i,v}^{(s)}$. We then replace all $g_{i,v}^{(s)}$'s with their lower bounds in \eqref{eq:update-obj-theta-gamma}, and optimize the new objective. One can show that the choice of $\xi^{(s)}_{j, v}$ as
\begin{equation*}
    \xi_{j,v}^{(s)} \defas \frac{\hat{\gamma}\tilde{x}_{j,v}}{(1-\hat{\gamma})\hat{\theta}_{v}^{(s)}+\hat{\gamma}\tilde{x}_{j,v}},
\end{equation*}
where $\hat{\btheta}^{(s)}$ and $\hat{\gamma}$ is the current estimate of $\btheta^{(s)}$ and $\gamma$, respectively, yields the closed-form updates
\begin{align*}
\theta_{v}^{(s)} & \propto \sum_i 
                   \delta_{s_i,s} \Big[\eta_{i0}x_{i,v} + \sum_{j<i}\eta_{ij}(1-\xi_{j,v}^{(s)})x_{j,v} \Big], \\
\gamma & = \frac{\sum_{i=1}^n\sum_{j<i}\sum_{v=1}^{V}\eta_{ij}x_{j,v}\xi_{j,v}^{(s_i)}}
                {\sum_{i=1}^n\sum_{j<i}\sum_{v=1}^{V}\eta_{ij}x_{i,v}}.
\end{align*}

\section{Related Work}
\label{sec:related_work}
In recent years there has been growing interest in developing computational methods to enhance the understanding of  real-world, social interactions within groups. 
\citet{DanescuNiculescuMizil:2011ja} developed a probabilistic framework to examine whether the phenomenon of linguistic accommodation holds for Twitter conversations. 
\citet{danescu2012echoes} studied the ability of the change of linguistic style markers to reveal social power difference in textual cascades.  
\citet{Blundell2012} proposed a non-parametric Bayesian model to infer latent groups from interaction data, which is further extended by \citet{Tan:2016ww} to allow time-varying receptivity of each person. 
Other works such as  \citet{Guo2015,Kawabata:2016ig,linderman2014discovering} focused on the problem of ``who influences whom?''---inferring the direct influence among individuals. 
Our work studies a novel question for textual conversation threads---``who started it?''---and provides a principled way to answer this question. 

On the other side, various methods, mostly non-parametric  \cite{Lewis2011,Zhou:2013vo,Bacry2012,Bacry2014,Xu2016icml,Hansen:2015dh,Reynaudbouret:2010kw,Lemonnier:2014jf} have been proposed to estimate the excitation kernel matrix for multivariate Hawkes processes. 
While these methods perform well empirically in estimating the trigger kernels, in the context of textual conversation threads, they can only answer the question of ``who influences whom?'', which is different from ours. 
Moreover, these methods only consider the unmarked MHPs; for textual conversation threads, however, one has to design an appropriate mark density to reflect subtle but critical linguistic adaptations, as what we do in our model.

\section{Experiments}
\label{sec:experiments}
This section aims to empirically evaluating the proposed two-stage approach for root source identification (see Sec.\,\ref{subsec:root-proba-overview}) by answering the following questions: 
(\textbf{Q1}) How accurate is the parameter estimation procedure? 
(\textbf{Q2}) Can the root source probabilities computed with the estimated MHP model reliably identify the root sources? 
(\textbf{Q3}) How does our method perform on real-world data \emph{with} the ground-truth reply structure? 
(\textbf{Q4}) How can our method provide insight into real-world conversation threads \emph{without} the ground truth? 
We shall answer \textbf{Q1} and \textbf{Q2} with experiments on a synthetic dataset in Sec.\,\ref{subsec:synthetic-data}, and then answer \textbf{Q2}--\textbf{Q4} with experiments on two real-world datasets in Sec.\,\ref{subsec:real-data}.

\subsection{Synthetic Data} 
\label{subsec:synthetic-data} 

\paragraph{Experimental Setup}
We consider a synthetic dataset generated by a marked MHP with $S=5$ sources. Every source $s$ has the same base intensity function with  $\bar{\mu}^{(s)}(t) = 1$ and $\rho_s = 0.1$, and the same exponential kernel $\kappa^{(s)}(t,t')=\frac{1}{\nu} \exp\{ -\frac{1}{\nu}(t' - t) \}$ with $\nu = 10$. 
The excitation matrix $\A$ is set to be symmetric with diagonal and off-diagonal entries $0.4$ and $0.1$, respectively, and the mark impact function $\beta(\x)$ is set to constant $1$. 
The five dimensions have average text lengths of $10,20,30,40$, and $50$, respectively; they also have a vocabulary inheritance rate of $\gamma=0.3$ and vocabulary parameters $\bm{\theta}^{(s)} \sim \Dir(\bm{1})$. The total vocabulary size is $V=5000$, and the total number of events is $n=10000$.

\myparagraph{Parameter Recovery (Q1)}

\begin{figure}[htbp]
  \centering
  \subfloat[Excitation matrix $\A$]{
    \includegraphics[width=.47\columnwidth]{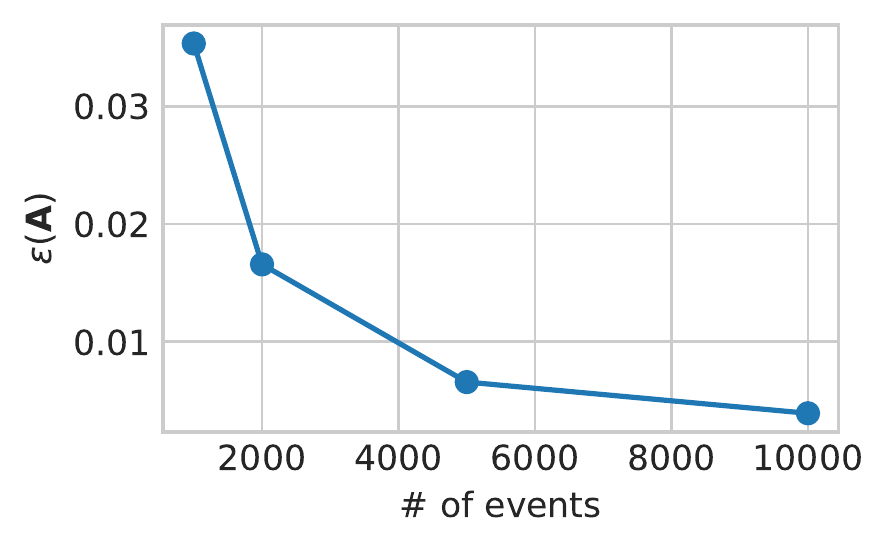} 
    \label{fig:rse-A}
  }
  \hfill
  \subfloat[Vocabulary parameter $\btheta^{(*)}$]{
    \includegraphics[width=.47\columnwidth]{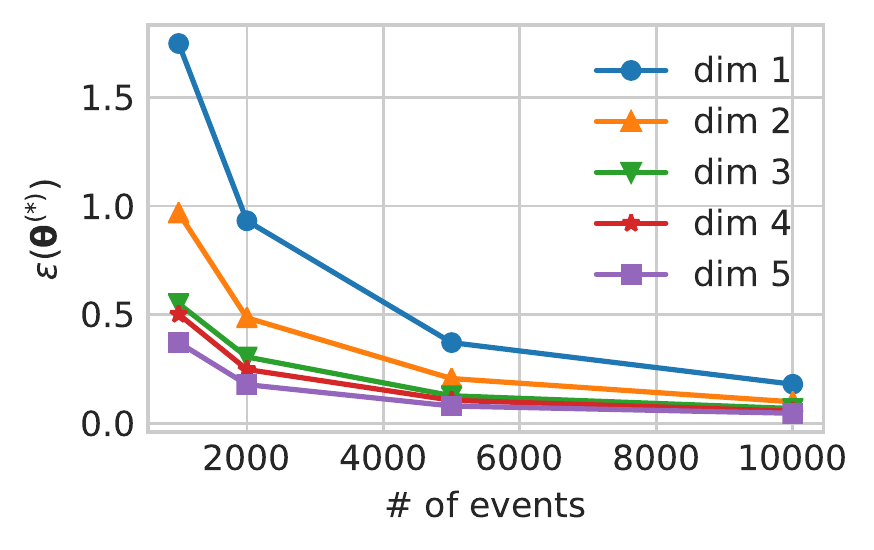} 
    \label{fig:rse-theta}
  }
  
  \caption{Relative error of the estimated parameters on the synthetic data with varying sample size}
  \label{fig:rse}
  \reducemargin
\end{figure}

First examine the reduction in the relative square errors (RSEs) for excitation matrix $\A$ and for vocabulary parameters $\btheta^{(s)}$'s when the number of events increases. 
As shown in Figure~\ref{fig:rse}, in both cases the RSEs decrease with more events observed, which verifies the effectiveness of the parameter estimation procedure. 
Overall, estimation for $\A$ is more accurate than that for $\btheta^{(s)}$'s, which is reasonable since the dimension of $\btheta^{(s)}$'s is much larger. 
Also, RSEs for $\btheta^{(s)}$'s are smaller for sources with longer text, suggesting that it is easier to recover the vocabulary parameters for vocal, loquacious individuals.
\vspace*{-0.3in}

\myparagraph{Root Source Identification (Q2)}

Now validate the ability of our method to identify root sources.  
To the best of our knowledge, no existing work provides a direct estimation of root source probabilities under the novel setting of this work; therefore, we validate the capacity of our method (\texttt{RP\_FIT}) via comparison with the following baseline families:
\begin{itemize}
  
  \item Heuristic running window baselines: use the normalized counts of comments from each source over the $M$ most recent events as an estimate of root source probabilities. Set $M=1, 10$, and $ \infty$ (denoted as \texttt{RW\_1}, \texttt{RW\_10}, and \texttt{RW\_inf}, respectively).

  \item Sub-model baselines: compute root source probabilities with temporal info only (\texttt{RP\_TEMP\_FIT}) and with textual info only (\texttt{RP\_MARK\_FIT}). That is, root source probabilities are calculated using the following simplified recursive equations: 
  \begin{equation*}
  \small
  \begin{aligned}
  \bar{r}_i^{(s)} & \propto\ \delta_{s_i, s} \mu^{(s)}(t_i) + \sum_{j<i} \bar{r}_j^{(s)} \lambda_j^{(s_i)}(t_i), \\
  \tilde{r}_i^{(s)} & \propto \delta_{s_i, s} f(\x_i|t_i,s_i) + \sum_{j<i} \tilde{r}_j^{(s)} f(\x_i|t_i,s_i,e_j).
  \end{aligned}
  \end{equation*}
  
\end{itemize}
Our method, as well as these baselines, are all compared with an oracle where the root source probabilities are computed using the true model parameters (\texttt{RP\_TRUE}). 

\begin{figure}[htbp]
  \centering
  \includegraphics[width=.95\columnwidth]{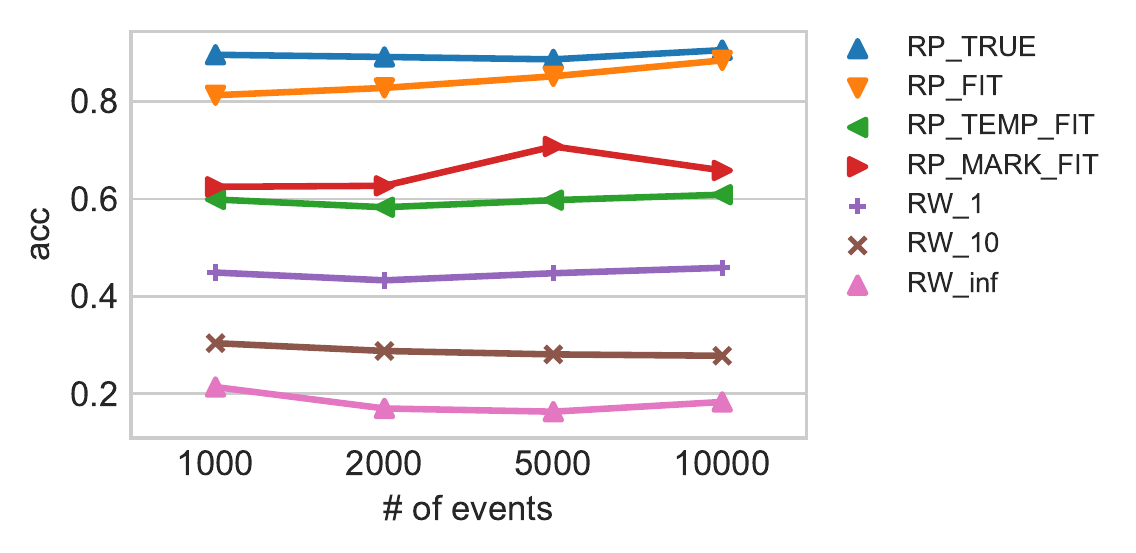}
  \caption{Accuracy of root source identification for each method on the synthetic data.}
  \label{fig:acc-root-source-identification}
  \reducemargin
\end{figure}

Figure~\ref{fig:acc-root-source-identification} shows the accuracy of the root sources identified by the different methods, compared against the true root sources (i.e., the root sources constructed by tracing the true branching structure). 
The \texttt{RP}-based methods outperform all other baselines in all cases, and as the sample size increases \texttt{RP\_FIT} converges to \texttt{RP\_TRUE}, the oracle method.

%

\subsection{Real Data}
\label{subsec:real-data}

\paragraph{Data}  Our model is also evaluated on the following two real-world datasets: 
\begin{itemize}
  \item \textbf{Reddit}: A collection of comments on 2016 US Election results, extracted from the website \url{reddit.com} \cite{redditcomment}.  Consider each post as an event and each user as a source. 
  \item \textbf{12 Angry Men:} A transcript of the 1957 legal-themed film, \emph{12 Angry Men}. Consider each juror as a source and each utterance as an event.  
\end{itemize}
We defer more details on data collection and processing to Appendix \ref{ap:subsec:data-collection-processing}.

One distinction between the two datasets is the availability of the ground-truth root sources.
Since the direct replying relationship between posts are observed for the \emph{Reddit} dataset, we are able to construct the ground-truth root sources and use them to \emph{quantitatively} evaluate model performance. 
The \emph{12 Angry Men} dataset---which exemplifies a more realistic and practical application of our method to real-world, off-line textual conversation threads---does not offer the replying structure for utterances; hence, we provide a comprehensive \emph{qualitative} analysis of our method on this dataset. 

\paragraph{Experimental Setup}
We choose the baseline intensity shape function as ${\bar{\mu}}^{(s)}(t) \equiv 1$ for both datasets, and excitation kernels $\kappa^{(s)}(t,t')=\frac{1}{\nu} \exp\{ -\frac{1}{\nu}(t' - t) \}$ with hyperparameters $\nu = 450$ and $\nu = 8$ for \emph{Reddit} and \emph{12 Angry Men}, respectively.  Other experimental setup is detailed in Appendix \ref{ap:susbsec:experimental-setup}. 

\paragraph{Evaluation Metrics}
For Reddit, we quantify model performance in root source identification with the following metrics: (a) total predicted log-probability of the true root sources, given by root source probabilities; (b) top-k accuracy, the proportion of true root sources ranked at top $k$ by root source probabilities ($k = 1, 10$).

\paragraph{Results on Reddit (Q2, Q3)}

Table~\ref{tab:reddit-results} compares the performance of \texttt{RP\_FIT} with the three aforementioned running window baselines. 
Our model, \texttt{RP\_FIT}, attains the best results on all three metrics
. The comparable Top-1 accuracy of \texttt{RW\_1} with \texttt{RP\_FIT} is simply due to the specificity of the dataset--a large number of posts are original comments, so the author is the root source--but for those comments with possibly different true root sources, \texttt{RP\_FIT} performs better.  
\begin{table}[htbp]
  \centering
  \resizebox{0.9\columnwidth}{!}{
    \begin{tabular}{lrrr}
\toprule
Method  & Log-Prob. & Top-1 Acc. & Top-10 Acc. \\ 
\midrule  
RW\_1 & -2300.28  &  0.74 & 0.77  \\
RW\_10 & -2807.71  & 0.11 & 0.77 \\
RW\_inf & -1807.84  & 0.04 & 0.30  \\
\midrule
RP\_FIT &\textbf{-852.15} & 0.74 & \textbf{0.79}  \\
\bottomrule
\end{tabular}
  }
  \caption{Model performance evaluated on \emph{Reddit}.}
  \label{tab:reddit-results}
  \reducemargin
\end{table}

\vspace*{-0.03in}
It is notable that \texttt{RP\_FIT} far outperforms \texttt{RW\_1} and \texttt{RW\_10} in total log-probability, because the baselines only consider sources that are temporally close, but our method also leverages textual information and accounts for the uncertainty in identification.

\myparagraph{Results on 12 Angry Men (Q4)}

As ground truth is not available for \emph{12 Angry Men}, we qualitatively analyze our model with a proxy task---inferring social power in conversation threads \cite{danescu2012echoes}. 
We hypothesize that the social power of an individual can be reflected by his/her ability to \emph{initiate conversational topics}; as a result, for individual $s$, the sum over the $s$-root probabilities for all events, i.e., $\sum_{i = 1}^n r^{(s)}_i$, may be a good measure for social power. 
We verify this hypothesis by ranking all jurors in the film using this measure and show the top five jurors in Table~\ref{tab:12-angry-men-power}. Juror 8 and Juror 3 are ranked as the top two, which exactly matches the film plot: the two jurors are the protagonist and antagonist, both heavily engaging in the discussion and frequently bringing up new talking points. 
The model also correctly ranks Juror 1 among the top, who serves as foreman and is responsible for maintaining order in the jury room. 
\begin{table}[htbp]
  \centering
  \resizebox{0.9\columnwidth}{!}{
    \begin{tabular}{clrl}
\toprule
Rank & Source & Power & Role\\ 
\midrule
1 & Juror 8 & 269.89 & insists acquittal\\ 
2 & Juror 3 & 53.34 & insists conviction\\ 
3 & Juror 7 & 40.29 \\ 
4 & Juror 1 & 36.99 & serves as foreman \\ 
5 & Juror 10 & 35.78 \\
\bottomrule
\end{tabular}
	
  }
  \caption{The top five most influential jurors in \emph{12 Angry Men} ranked by the root source probability measure.}
  \label{tab:12-angry-men-power}
  \reducemargin
\end{table}

\begin{figure}[htbp]
  \centering    
  \resizebox{\columnwidth}{!}{
    \begin{tikzpicture}[
	scale=1,
    axis/.style={thick, ->, -latex},
    utterance/.style={text width=30ex, align=left, node font=\bf},
    ]
    
  \node[draw, node font=\normalsize] (J8) at (0, 0) {\bf Juror 8};    
	\node[draw, node font=\normalsize] (J12) at (7, 0) {\bf Juror 12};  
	
	\node[below=0.4 of J8, utterance] (a11) {Actually those two \textcolor{cGreen} {witnesses} were the entire case for the prosecution. \textcolor{cRed}{Supposing they are wrong}... };
	\node[below=0.4 of a11, utterance] (a12) {\textcolor{cRed}{Could they be wrong}? };
	\node[below=0.4 of a12, utterance] (a13) {They are only \textcolor{cBlue}{people}. \textcolor{cBlue}{People} make mistakes. };
	\node[below=0.4 of a13, utterance] (a14) {\textcolor{cRed}{Could they be wrong}?};
	
	\node[below=1.4 of J12, utterance] (a21) {What do you mean \textcolor{cRed}{supposing they're wrong}? What's the point of having \textcolor{cGreen}{witness} at all?};
	\node[below=0.4 of a21, utterance] (a22) {Those \textcolor{cBlue}{people} sat on the stand under oath!};
	\node[below=0.8 of a22, utterance] (a23) {Well, no. I don't think so.};
	
	\draw[axis] (a11) -- (a21);
	\draw[axis] (a21) -- (a12);
	\draw[axis] (a12) -- (a22);
	\draw[axis] (a12.west) to[bend right] (a14.west);
	\draw[axis] (a22) -- (a13);
	\draw[axis] (a14) -- (a23);
\end{tikzpicture}
  }
  \caption{Vocabulary inheritance in \textbf{12 Angry Men}.}
  \label{fig:12-angry-men-word-inheritance}
  \reducemargin
\end{figure}
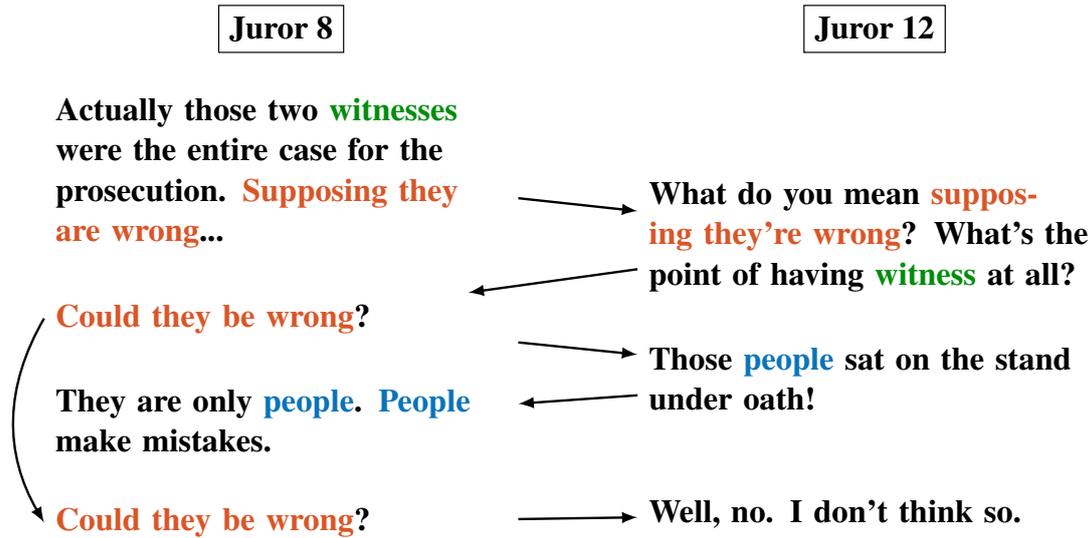

We also investigate the ``mini-conversations'' found by our model; each mini-conversation is defined as a collection of the comments rooted by a comment that initiates a new topic and is constructed by treating the learned variational variables $\eta_i$'s in \eqref{eq:Q} as a proxy of the branching structure. 
We observe that many of these mini-conversations agree with human intuition and exhibit clear vocabulary inheritance. In the mini-conversation shown in Figure \ref{fig:12-angry-men-word-inheritance}, for example, the word ``witness'', the phrase ``could they be wrong'' and the word ``people'' are repeated, which verifies the model assumption that a response comment may inherit words from the comment it replies to. 

\section{Conclusion}
\label{sec:conclusion}
We address the problem of  identifying root sources in textual conversation threads. 
We propose a marked multivariate Hawkes process model to describe the dynamics of textual cascades, and then introduce a novel concept, \emph{root source probability}, to quantify the uncertainty of identified root sources. 
An efficient, dynamic-programming-based algorithm is derived to compute root source probabilities, and a parameter estimation procedure based on variational inference is developed.
Experiments on synthetic and real-world datasets show that the proposed method can identify root sources that agree with both ground truth and human intuition.

\newpage

\appendix
\newpage
\onecolumn

%

\section{Technical Details}

\subsection{Graphical Model for the Proposed Marked MHP}
\label{ap:subsec:graphical-model}

\begin{figure}[H]
	\centering
	\includegraphics[width=0.5\columnwidth]{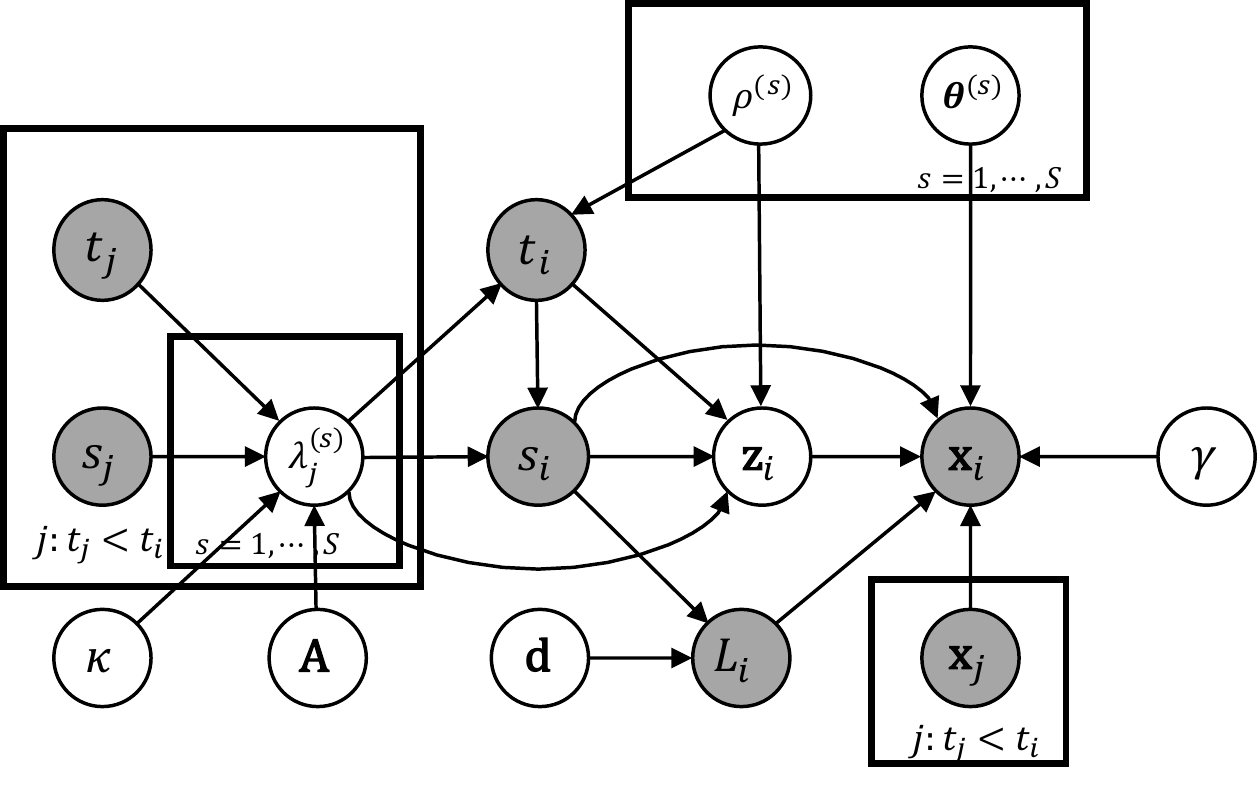}
	\caption{The graphical model for the proposed marked MHP for modeling group conversation}
\end{figure}

\subsection{Proof of Proposition \ref{prop:rooted-probability-calculation}}
\label{proof:rooted-probability-calculation}

\begin{proof}
  First, the space $\cZ_i^{(s)}([n])$ can be rewritten as the Cartesian product of $\cZ_i^{(s)}([i])$ and $\cZ(\{i+1,\ldots, n\})$. Given the conditional independence of $\z_{[i]}$ and $\z_{\{i + 1, \ldots n\}}$ for any $i$, we marginalize the latter term from \eqref{eq:rooted-probability-definition}, yielding 
  \begin{equation}
  r_i^{(s)} =\sum_{\z_{[i]} \in \cZ_i^{(s)}([i])} P(\z_{[i]}|\cH_{t_i}).
  \label{eq:rooted-probability-calculation-simplified}
  \end{equation}
  Note that $\z_{[i]}\in\cZ_i^{(s)}([i])$ belongs to exactly one of the following two cases:
  \begin{enumerate}
    \item $z_{i0}=1$ and $s_i=i$.
    \item $z_{ij}=1$ and $\z_{[i-1]}\in\cZ_j^{(s)}([i-1])$ for some $j < i$.
  \end{enumerate}
  Thus the right hand side of  \eqref{eq:rooted-probability-calculation-simplified} can be rewritten as
  \begin{equation*}
  \begin{aligned}
  & \sum_{ \z_{[i-1]} \in \cZ([i-1])} \delta_{s_i,s} 
  P(\z_{[i-1]}|\cH_{t_i-})P(z_{i0}=1|\cH_{t_i})  \\
  & \qquad + \sum_{j<i} \sum_{ \z_{[i-1]} \in \cZ_j^{(s)}([i-1]) } P(\z_{[i-1]}|\cH_{t_i-})P(z_{ij}=1|\cH_{t_i}).
  \end{aligned}
  \end{equation*}
  Since $\sum_{\z_{[i-1]}\in\cZ_j^{(s)}([i-1])}P(\z_{[i-1]}|\cH_{t_i-})$
  is indeed $r_j^{(s)}$, we have 
  \begin{equation*}
  r_i^{(s)} = \delta_{s_i,s}P(z_{i0}=1|\cH_{t_i}) + \sum_{j<i}r_j^{(s)}P(z_{ij}=1|\cH_{t_i}).
  \end{equation*}
  Finally, since
  \begin{align*}
  P(\z_i|\cH_{t_i}) &= P(\z_i|\t_i, \s_i, \x_i, \cH_{t_i-})   \\
  & \propto P(\z_i|t_i,s_i,\cH_{t_i-}) P(\x_i|\z_i,t_i,s_i,\cH_{t_i-}),
  \end{align*}
  with \eqref{eq:P(z_e)}--\eqref{eq:mark-density} we have
  \begin{align*}
  r_i^{(s)} \propto~ & \delta_{s_i,s}P(z_{i0}=1|t_i,s_i,\cH_{t_i-})f(\x_i|t_i,s_i)  + \sum_{j<i}r_j^{(s)}   P(z_{ij}=1|t_i,s_i,\cH_{t_i}) f(\x_i|t_i,s_i,e_j) \\
  \propto~ & \delta_{s_i,s}\mu^{(s_i)}(t_i)f(\x_i|t_i,s_i)  
  +\sum_{j<i}r_j^{(s)}\lambda_j^{(s_i)}(t_i)f(\x_i|t_i,s_i,e_j).
  \end{align*}
  \normalsize
  
\end{proof}

\subsection{The Variational Lower Bound $\tilde{\cL}(\bTheta, Q)$}
\label{ap:subsec:ELBO}

The complete likelihood is
\begin{align*}
 P(\cH_T,\z_{[n]}|\Theta) 
= &  \exp\left(-\int_{0}^T\sum_s\lambda^{(s)}(t|\cH_{t-})\,dt\right) \prod_{i=1}^n \lambda^{(s_i)}(t_i|\cH_{t_i-}) \\
  & \times \prod_{i=1}^n \left[ \left(\frac{\mu^{(s_i)}(t_i)}{\lambda^{(s_i)}(t_i|\cH_{t_i-})}f(\x_i|t_i,s_i)\right)^{z_{i0}}
   \times\prod_{j<i}\left(\frac{\lambda_j^{(s_i)}(t_i)}{\lambda^{(s_i)}(t_i|\cH_{t_i-})}f(\x_i|t_i,s_i, e_j)\right)^{z_{ij}}\right].
\end{align*}
Then the complete log likelihood with the parameterization specified in the paper becomes:
\begin{align*}
& \log P(\cH_T,\z_{[n]}|\bTheta) \\
= & - \sum_s  \rho_s \int_{0}^T \bar{\mu}^{(s)}(t)\,dt  - \sum_s \sum_{i=1}^n \alpha_{s,s_i} \beta(\x_i) \int_{t_i}^T \kappa(t_i,t)\,dt \\
& + \sum_{i=1}^n z_{i0} \log \left[ \rho_{s_i} \bar{\mu}^{(s_i)}(t_i) f(\x_i|t_i,s_i) \right] 
 + \sum_{i=1}^n \sum_{j<i}z_{ij} \log \left[\alpha_{s_i, s_j} \kappa(t_j,t_i) f(\x_i|t_i,s_i,e_j) \right] .
\end{align*}
Therefore the variational lower bound $\tilde{\cL}(\Theta,Q)$ is 
\begin{align*}
\tilde{\cL}(\Theta,Q) \defas & \bbE_{Q} \left[\log P(\cH_T,\z_{[n]}|\bTheta)\right] - \bbE_{Q} \left[ \log Q(\z_{[n]}) \right] \\
= & -\sum_s  \rho_s \int_{0}^T \bar{\mu}^{(s)}(t)\,dt  
    - \sum_s\sum_{i=1}^n \alpha_{s,s_i} \beta(\x_i) \int_{t_i}^T \kappa(t_i,t)\,dt \\
& +\sum_{i=1}^n\eta_{i0}\log\left[\rho_{s_i}\bar{\mu}^{(s_i)}(t_i)f(\x_i|t_i,s_i)\right]\\
& +\sum_{i=1}^n\sum_{j<i}\eta_{ij}\log\left[\alpha_{s_i,s_j}\kappa(t_j,t_i)f(\x_i|t_i,s_i,e_j)\right] 
  -\sum_{i=1}^n\Big(\eta_{i0}\log\eta_{i0}+\sum_{j<i}\eta_{ij}\log\eta_{ij}\Big) .
\end{align*}

\subsection{Derivation of Updates for $\brho$ and $\A$ with Empirical Bayes Priors}
\label{ap:subsec:update-details-EB}

The modified lower bound using Gamma priors $\rho_s \sim \Gam(a_{\rho}^{(s)}, b_{\rho})$ and $\alpha_{s,s'} \sim \Gam(a_{\alpha}^{(s)},b_{\alpha})$ is
\begin{align*}
& \tilde{\cL}(\Theta,Q)\\
= & -\sum_s  \rho_s \int_{0}^T \bar{\mu}^{(s)}(t)\,dt 
    - \sum_s\sum_{i=1}^n \alpha_{s,s_i} \beta(\x_i) \int_{t_i}^T \kappa(t_i,t)\,dt \\
& + \sum_{i=1}^n\eta_{i0}\log\left[\rho_{s_i}\bar{\mu}^{(s_i)}(t_i)f(\x_i|t_i,s_i)\right]
  +\sum_{i=1}^n\sum_{j<i}\eta_{ij}\log\left[\alpha_{s_i,s_j}\kappa(t_j,t_i)f(\x_i|t_i,s_i,e_j)\right]\\
& -\sum_{i=1}^n\left(\eta_{i0}\log\eta_{i0}+\sum_{j<i}\eta_{ij}\log\eta_{ij}\right) \\
& + \sum_s \left[(a_{\rho}^{(s)} - 1)\log{\rho_s}-b_{\rho}\rho_s \right] 
  + \sum_s \sum_{s'} \left[(a_{\alpha}^{(s)} - 1)\log{\alpha_{s,s'}}-b_{\alpha}\alpha_{s,s'}\right].
\end{align*}
Taking first order derivatives regarding $\rho_s$ and $\alpha_{s,s'}$ gives the modified updates
\begin{align*}
\rho_s & =\frac{a_{\rho}^{(s)} - 1 + \sum_{i=1}^n \delta_{s_i,s}\eta_{i0}}{b_{\rho}+\int_{0}^T\bar{\mu}^{(s)}(t)\,dt},\\
\alpha_{s,s'} & = \frac{a_{\alpha}^{(s)} - 1 +\sum_{i=1}^n\sum_{j<i}\delta_{s_i,s} \delta_{s_j,s'}\eta_{ij} }
{b_{\alpha} + \sum_{i=1}^n \delta_{s_i, s'}\beta(\x_i)\int_{t_i}^T\kappa^{(s_i)}(t_i,t)\,dt }  .
\end{align*}

\subsection{Derivation of Updates for $\bTheta$ and $\gamma$}
\label{ap:subsec:update-details-vocabulary-param}

For any coefficient $\xi_{j,v}^{(s)}\in(0,1)$, applying Jensen's inequality, we have
\begin{equation}
\log\left[ (1-\gamma)\theta_{v}^{(s)}+\gamma\tilde{x}_{j,v} \right]  
\geq  (1-\xi_{j,v}^{(s)}) \log \frac{(1-\gamma)\theta_{v}^{(s)}}{1-\xi_{j,v}^{(s)}} 
+ \xi_{j,v}^{(s)} \log \frac{\gamma\tilde{x}_{j,v}}{\xi_{j,v}^{(s)}} ,
\end{equation}
which leads to a lower bound of $g_{i,v}^{(s)}(\btheta,\gamma)$ as
\begin{equation}
\begin{aligned}
 g_{i,v}^{(s)}  (\btheta,\gamma) 
& \geq  \eta_{i0} x_{i,v} \log\theta_{v}^{(s)}  
+ \sum_{j<i} \eta_{ij}x_{i,v} \left[ (1-\xi_{j,v}^{(s)}) \log \frac{(1-\gamma)\theta_{v}^{(s)}}{1-\xi_{j,v}^{(s)}} 
+ \xi_{j,v}^{(s)} \log \frac{\gamma\tilde{x}_{j,v}}{\xi_{j,v}^{(s)}} \right] \\
& =  
\Big[ \eta_{i0}x_{i,v}+\sum_{j<i}\eta_{ij}x_{i,v}(1-\xi_{j,v}^{(s)}) \Big] \log\theta_{v}^{(s)}  \\
& \quad  + \sum_{j<i}\eta_{ij}x_{i,v}[(1-\xi_{j,v}^{(s)}) \log(1-\gamma)+\xi_{j,v}^{(s)}\log\gamma]  
+  \mathrm{const}.
\end{aligned}
\label{eq:g-lower-bound}
\end{equation}
Define 
\begin{equation*}
  \tilde{g}^{(s)}_{j, v} \defas  
  \Big[ \eta_{i0}x_{i,v}+\sum_{j<i}\eta_{ij}x_{i,v}(1-\xi_{j,v}^{(s)}) \Big] \log\theta_{v}^{(s)}  \\
   + \sum_{j<i}\eta_{ij}x_{i,v}[(1-\xi_{j,v}^{(s)}) \log(1-\gamma)+\xi_{j,v}^{(s)}\log\gamma].
\end{equation*}
Solving the optimization problem
\begin{equation*} 
\begin{aligned}
\max_{\btheta,\gamma} & \sum_{s=1}^{S}\sum_{i=1}^n\sum_{v=1}^{V}\delta_{s_i,s} \tilde{g}_{i,v}^{(s)}(\btheta,\gamma), \\
& \text{s.t.} \sum_{v=1}^{V}\theta_{v}^{(s)}=1, \forall s \in [S], 
\end{aligned}
\end{equation*}
leads to the updates
\begin{align*}
\theta_{v}^{(s)} & \propto \sum_i \delta_{s_i,s} [\eta_{i0}x_{i,v} + \sum_{j<i}\eta_{ij}(1-\xi^{(s)}_{j,v}) x_{i,v}], \\
\gamma & = \frac{\sum_{i=1}^n\sum_{j<i}\sum_{v=1}^{V}\eta_{ij}x_{i,v}\xi_{j,v}^{(s_i)}}
{\sum_{i=1}^n\sum_{j<i}\sum_{v=1}^{V}\eta_{ij}x_{i,v}}.
\end{align*}

We choose $\xi_{j,v}^{(s)}$ to be
\begin{equation}
\xi_{j,v}^{(s)}=\frac{\hat{\gamma}\tilde{x}_{j,v}}{(1-\gamma)\hat{\theta}_{v}^{(s)}+\hat{\gamma}\tilde{x}_{j,v}},
\end{equation}
where $\hat{\btheta}^{(s)}$ and $\hat{\gamma}$ are the current estimate of $\btheta^{(s)}$ and $\gamma$, respectively. This is because this choice makes the inequality in \eqref{eq:g-lower-bound} tight with the old estimates and thus guarantees that the objective function increases monotonically.

\section{Experimental Details}

\subsection{Data Collection and Processing}
\label{ap:subsec:data-collection-processing}

\paragraph{Reddit}

Our Reddit data are constructed from the data dump provided by the website  \citet{redditcomment}.
We consider a very popular article, on the \emph{politics} subreddit, titled, ``2016 Election Day Returns Megathread'', and extract all comments occurring on the thread from November 8 through November 15 of 2016.
This interval begins when the thread opened, on Election Tuesday.  
We then filter out the authors who commented less than $5$ times.  
Comments that respond to deleted comments are also discarded.
As all the comments have already organized in tree hierarchies, we define the root source (as known as O.P) of each comment to be the user of its first level parent comment.

\paragraph{12 Angry Men}

The transcript of the film \emph{12 Angry Men} is obtained from \url{https://github.com/richardkwo/bayesian-echo-chamber/tree/master/data}. We label each juror with his juror ID in the film. 

\subsection{Detailed Experimental setup}
\label{ap:susbsec:experimental-setup}

\paragraph{Empirical Bayes} 

We set the hyperparameters $a_{\rho}^{(s)} = N_s$, $b_{\rho} = T/c$, $a_{\alpha}^{(s)} = N_s$, $b_{\alpha} = T/(1-c)$. 
Here $N_s$ is the total number of events on source $s$, and $c$ is the expected proportion of baseline events. 
We choose $c=1/10$ to encourage original posts. 

\subsection{Inference Algorithm Scalability}

\begin{figure}[htbp]
	\centering
	\includegraphics[width=0.5\columnwidth]{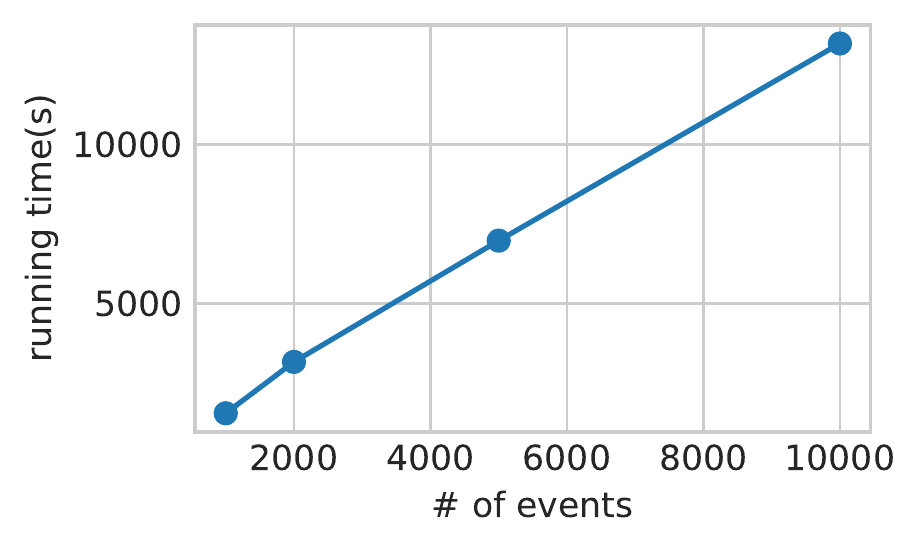}
	\caption{Training time scales linearly with the number of events.}
	\label{fig: running time}
\end{figure}

Algorithm running time is recorded in the synthetic data experiments. As shown in Figure~\ref{fig: running time}, model training time scales linearly with the number of events, support our claim of the efficiency of the proposed dynamic programming algorithm for computing all the root source probabilities. 

\subsection{Qualitative Analysis on Reddit}

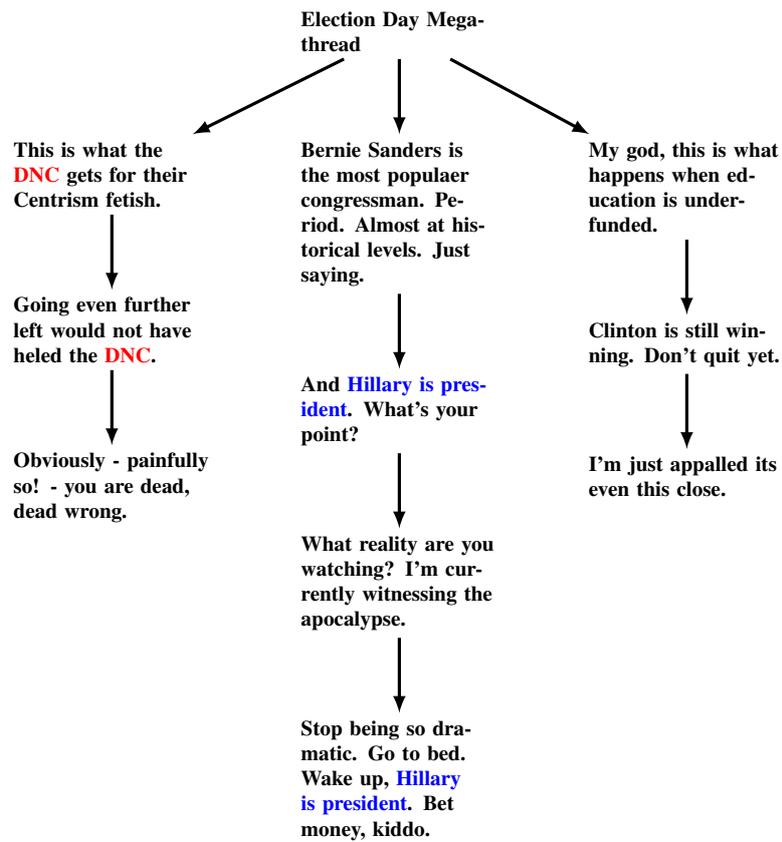
\begin{figure}[htbp]
\centering
\resizebox{0.7\textwidth}{!}{\begin{tikzpicture}[
	scale=1,
    axis/.style={very thick, ->, -latex},
    post/.style={text width=20ex, node font=\bf \scriptsize}
    ]
    
    \node[post] (R) at (0, 0) {Election Day Mega-thread};
	\node[post, below left=of R] (p11) {This is what the {\color{red} DNC} gets for their Centrism fetish.};
	\node[post, below=of p11]  (p12) {Going even further left would not have heled the {\color{red} DNC}.};
	\node[post, below=of p12]  (p13) {Obviously - painfully so! - you are dead, dead wrong.};
	\node[post, below=of R]    (p21) {Bernie Sanders is the most populaer congressman. Period. Almost at historical levels. Just saying.};
	\node[post, below=of p21]  (p22) {And {\color{blue} Hillary is president}. What's your point?};
	\node[post, below=of p22]  (p23) {What reality are you watching? I'm currently witnessing the apocalypse.};
    \node[post, below=of p23]  (p24) {Stop being so dramatic. Go to bed. Wake up, {\color{blue} Hillary is president}. Bet money, kiddo.};
    \node[post, below right=of R] (p31) {My god, this is what happens when education is underfunded.};
	\node[post, below=of p31]  (p32) {Clinton is still winning. Don't quit yet.};
	\node[post, below=of p32]  (p33) {I'm just appalled its even this close.};
	
	\draw[axis] (R) -- (p11);
	\draw[axis] (p11) -- (p12);
	\draw[axis] (p12) -- (p13);
	\draw[axis] (R) -- (p21);
	\draw[axis] (p21) -- (p22);
    \draw[axis] (p22) -- (p23);
    \draw[axis] (p23) -- (p24);
    \draw[axis] (R) -- (p31);
	\draw[axis] (p31) -- (p32);
	\draw[axis] (p32) -- (p33);
	
\end{tikzpicture}}
\caption{Vocabulary inheritance in \textbf{Reddit}. Note inheritance of the word ``DNC'' in the first branch and the statement ``Hillary is president'' in the second branch.}
\label{fig:comment_tree_reddit}
\end{figure}

We also provide qualitative analysis on \textbf{Reddit} data similar to that on \textbf{12 Angry Men}. 
Table~\ref{tab:reddit_power} shows the top five most influential users in \textbf{Reddit} ranked by the root source probability measure. Our power measure aligns well with the rankings by \emph{Reddit Gold}, a virtual coin awarded by other users to valued comments. 
Figure~\ref{fig:comment_tree_reddit} shows incidences of vocabulary inheritance in the text of \emph{Reddit} comments. We visualize comments of which the root sources are correctly identified by our model. Note inheritance of the word ``DNC'' in the first branch and the statement ``Hillary is president'' in the second branch.

\begin{table}[htbp]
\centering
\begin{tabular}{clrcc}
\toprule
Rank & Source & Power & Gold & Gold Rank \\ 
\midrule
1 & User E & 15.94 & 385 &   2 \\ 
2 & User S1 & 11.66 &  72 &  17 \\ 
3 & User S2 & 10.38 &  44 &  28 \\ 
4 & User S3 & 10.32 &  73 &  16 \\ 
5 & User R & 10.20 &  73 &  15 \\
\bottomrule
\end{tabular}
\label{tab:influential_sources_reddit}
\caption{The top five most influential users in \emph{Reddit} ranked by the root source probability measure.}
\label{tab:reddit_power}
\end{table}

\end{document}